\documentstyle[aps,prl,multicol,psfig]{revtex}
\begin{document}
\draft
\title{Phys. Rev. E, Vol. 54, R4560-4563 (1996)\\[1cm]
Swirling Granular Matter: From Rotation to Reptation}
\author{Michael A. Scherer$^1$, Volkhard Buchholtz$^2$, Thorsten P\"oschel$^2$,
 and Ingo Rehberg$^1$}
\address{$^1$ Institut f\"ur Experimentelle Physik,
          Otto-von-Guericke-Universit\"at,\\
          Postfach 4120,
          D-39016 Magdeburg,
          Germany}
\address{$^2$ Institut f\"ur Physik,
          Humboldt-Universit\"at zu Berlin, 
          Invalidenstra\ss e 110,\\ D-10115 Berlin,
          Germany.}
          
\date{\today}
\maketitle
\begin{abstract}
A novel effect in a granular material under swirling motion of the container 
is presented. At low packing densities the material rotates in the same
direction as the swirling motion of the container (rotation). At higher
densities the cluster of granular material rotates in opposite direction
(reptation). The change of the direction of the motion of the cluster takes
place at a critical packing density while the diffusion coefficient changes
significantly. The measured critical density of the packing is in good
agreement with results obtained by molecular dynamics simulation.
\end{abstract}
\begin{multicols}{2}

The fascination of the flow behavior of granular material like sand is
due to the fact that under certain circumstances it behaves like a
fluid -- an hour glass is the most popular example. Its solid-like behavior 
is obvious as well: A pile of sand is stable and it can undergo plastic
deformation, but, unlike a fluid, it does not dissolve under gravity. The
exotic behavior of granular matter and the sometimes unexpected observations
have stimulated many scientists to focus their scientific interest to sand-
like materials (see e.g.~\cite{Knight,Ippolito,Melo}). For a recent review see~\cite{Hayakawa}.

We present a novel effect which demonstrates to a certain amount both
the similarity and the dissimilarity between granular material and
fluid. The experiment is extremely robust, easily visualizable, and
can be performed even as a kitchen table experiment using a bunch of
marbles and a pot. The basic idea is to bring the pot in a swirling
motion, the motion one frequently uses to stir up the bouquet of a
glass of wine. If the pot contains only one or a few spheres, they
will follow the rotation of the swirl in a similar fashion as the wine in
the glass does. When one successively increases the number of spheres the
angular velocity of the cluster decreases with increasing density.
At a certain critical density an intriguing effect appears: The angular
velocity becomes negative, i.e. the cluster rotates in the opposite direction.
Now the collective behavior is more reminiscent of a pancake rotated by
the swirling motion of a frying pan. The first regime of motion we
call ``rotation'', the second one ``reptation''. In this paper we
present the first qualitative and quantitative description of this
crossover effect. The experiments are accompanied by a two-dimensional
molecular dynamics simulation. The numerical results are qualitatively
and quantitatively in good agreement with the experimental findings.
 
\begin{figure}[htbp]
\begin{minipage}{8 cm}
\centerline{\psfig{figure=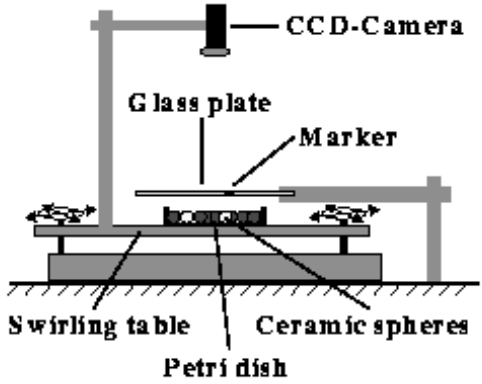,width=7cm,angle=0}}
\caption {Experimental setup.}
\label{aufbau}
\end{minipage}
\end{figure}

The experimental setup is shown in fig.~\ref{aufbau}. A Petri dish of
inner diameter $9.0~cm$ is mounted on a table performing a circular
vibration, i.e. each point of the table moves along a circular line
during one oscillation period. For the experiments presented here, its
frequency is chosen to be $2.5~s^{-1}$ and the amplitude of the
swirling motion is $1.27~cm$. A camera is fixed on the oscillating
table, thus the analysis can be done in a comoving frame by visual
inspection of the image displayed on a monitor. In between the Petri
dish and the camera there is a glass plate with a marker which is
fixed in the laboratory frame. Thus the position of the marker
relative to the dish indicates the direction of the momentary
acceleration. The Petri dish is partially filled with a monolayer of
ceramic spheres (density $1.74 g/cm^3$) of diameter $1.223 \pm 0.035
cm$. The number of particles ranges between 1 and 42, where 42
corresponds to the closest packing of spheres. We want to focus here
only on the density dependence of the effect. The dependence on the
amplitude, the frequency, the material properties etc. will be
discussed in more detail in a forthcoming paper.

The system described above was simulated using two-dimensional molecular 
dynamics. We applied the soft particle Ansatz by Cundall and
Strack~\cite{HaffWerner:1986} including interaction between colliding
particles in normal and tangential direction. Two colliding particles
$i$ and $j$ feel the force
\begin{equation}
  \label{force}
\vec{F}_{ij} = F_{ij}^N\, \vec{n} + F_{ij}^T\, \vec{t}
\label{force1}
\end{equation}
with 
\begin{eqnarray}
  F_{ij}^N &=& Y \cdot \left(R_i+R_j- \left| \vec{r}_i-\vec{r}_j\right| 
  \right) \nonumber \\
&& ~~~- m_{ij}^{eff} \cdot \gamma_N \cdot 
  (\dot{\vec{r}}_i - \dot{\vec{r}}_j) \cdot\vec{n}
\label{fnormal}\\ 
F_{ij}^T &=& \mbox{sign} (v_{ij}^{rel}) \cdot \min\left( m_{ij}^{eff} \gamma_T \left|
v_{ij}^{rel}\right| , \mu \left| F_{ij}^{N}\right| \right) 
\label{ftang} \\
v_{ij}^{rel} &=& (\dot{\vec{r}}_i - \dot{\vec{r}}_j)  \cdot
\vec{t}+ R_i \cdot \Omega_i + R_j \cdot \Omega_j 
\label{surfvelocity}\\ 
m_{ij}^{eff} &=& \frac{m_i \cdot m_j}{m_i + m_j} ~.
\label{meff}
\end{eqnarray}
$Y=8\cdot 10^6~g~cm^2$ is the Young modulus, $\gamma_N=800~s^{-1}$
and $\gamma_T=3000~s^{-1}$ are the damping coefficients in normal
and tangential direction and $\mu=0.5$ is the Coulomb friction
coefficient. The unit vectors in normal and tangential direction are given by
\begin{eqnarray}
\vec{n} &=& \frac{\vec{r}_i-\vec{r}_j}{\left|\vec{r}_i-\vec{r}_j\right|}\\
\vec{t} &=& \left( {0 ~ -1}\atop{1 ~~~~ 0}  \right) \cdot \frac{\vec r_i -\vec r_j }{\left| \vec r_i-\vec r_j \right|}~.
\end{eqnarray}

The parameters are chosen to represent the material properties of the
ceramic spheres used in the experiment as close as possible.
Eq.~(\ref{surfvelocity}) describes the relative velocity of the
surfaces of the particles at the point of contact and eq.~(\ref{meff})
gives the effective mass.  Eq.~(\ref{ftang}) takes the Coulomb
friction law into account, saying that two particles slide on top of
each other if the shear force overcomes $\mu$ times the normal force.
For the integration of Newton's equations we applied a Gear predictor
corrector method of fifth order~\cite{AllenTildesley:1987}.

\begin{figure}[htbp]
\begin{minipage}{8 cm}
\centerline{\psfig{figure=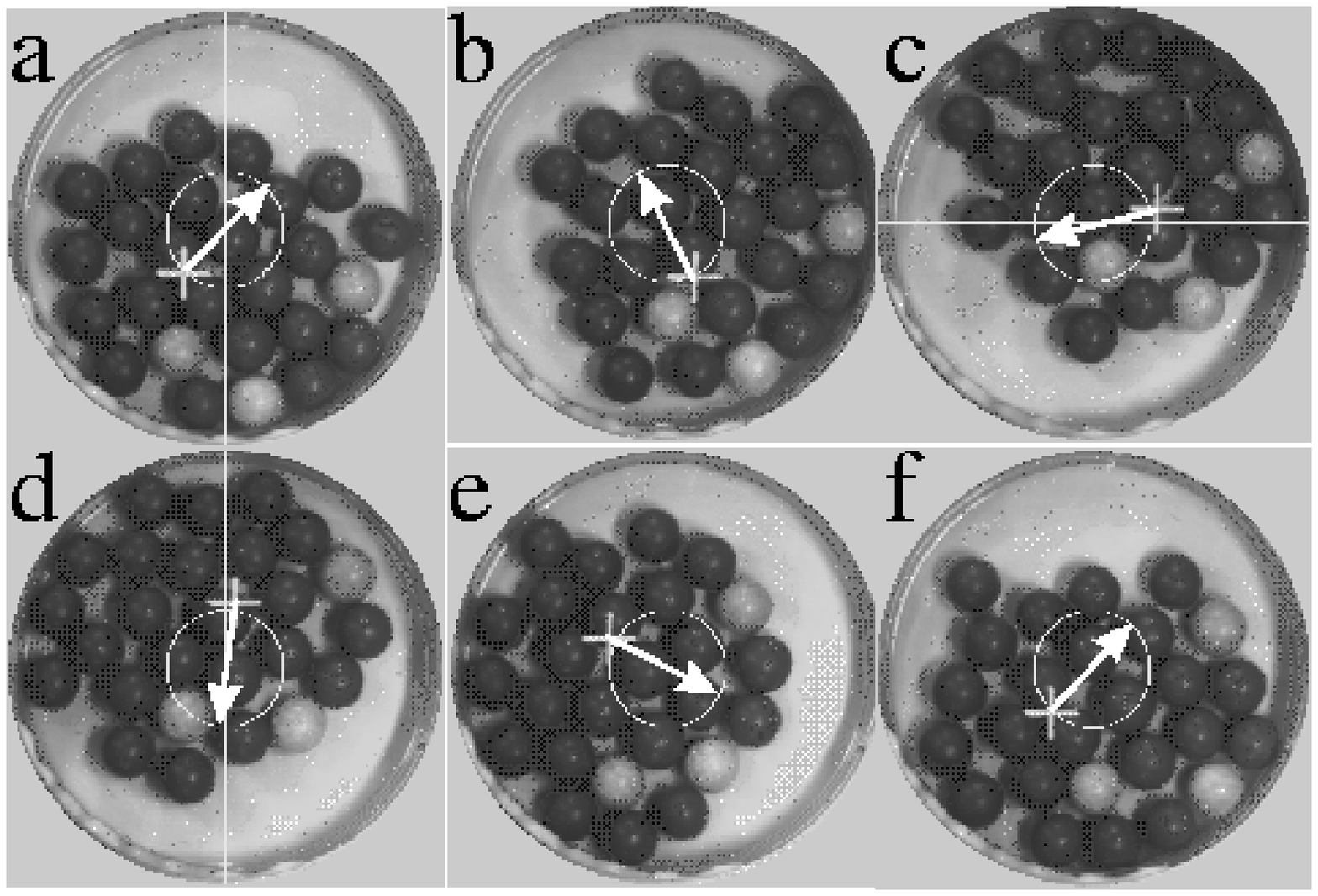,width=7cm,angle=0}}
\psfull
\centerline{\psfig{figure=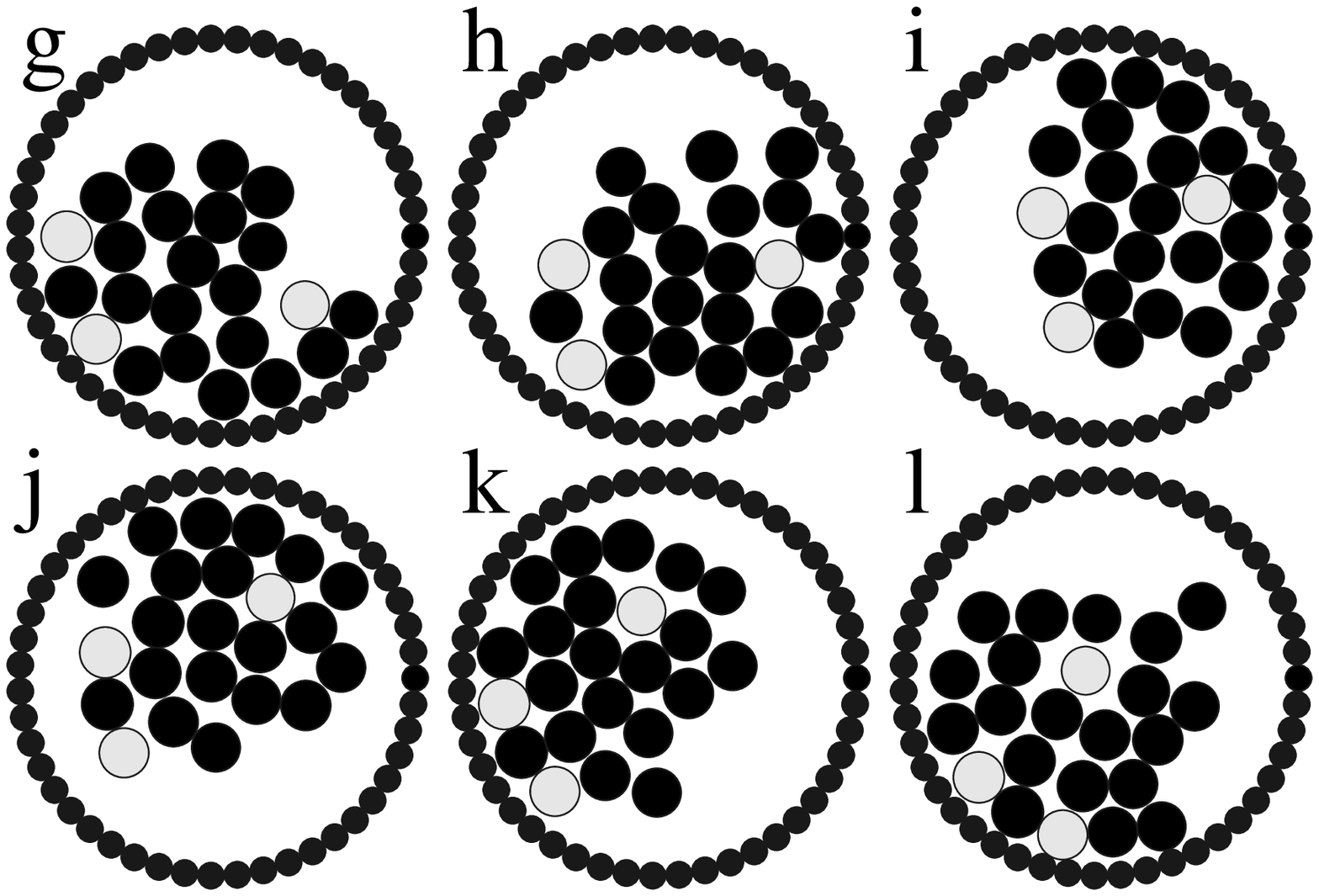,width=7cm,angle=0}}
\caption {Temporal behavior of a system of 26 spheres
  during one counterclockwise cycle of the swirl. The arrow shows the
direction of the
  momentary acceleration. The lower 6 figures show the corresponding
  molecular dynamics simulation.}
\label{MovieLow}
\end{minipage}
\end{figure}

The time series of images in fig.~\ref{MovieLow} visualizes the motion
of 26 spheres\cite{BlackAndWhite} in a
counterclockwise rotating Petri dish at low packing density. They have
been taken at time intervals $T/5$, with $T$ being the period of the
driving motion of the Petri dish. The evolution of the positions of
the white spheres in figs.~\ref{MovieLow}a to~\ref{MovieLow}f indicates
that the cluster of spheres also rotates in counterclockwise direction.
Figs.~\ref{MovieLow}$g$ to~\ref{MovieLow}$l$ show corresponding
to figs.~\ref{MovieLow}$a$ to~\ref{MovieLow}$f$ snapshots from a molecular
dynamics simulation using the same parameters as the experiment.
\begin{figure}[htbp]
\begin{minipage}{8 cm}
\centerline{\psfig{figure=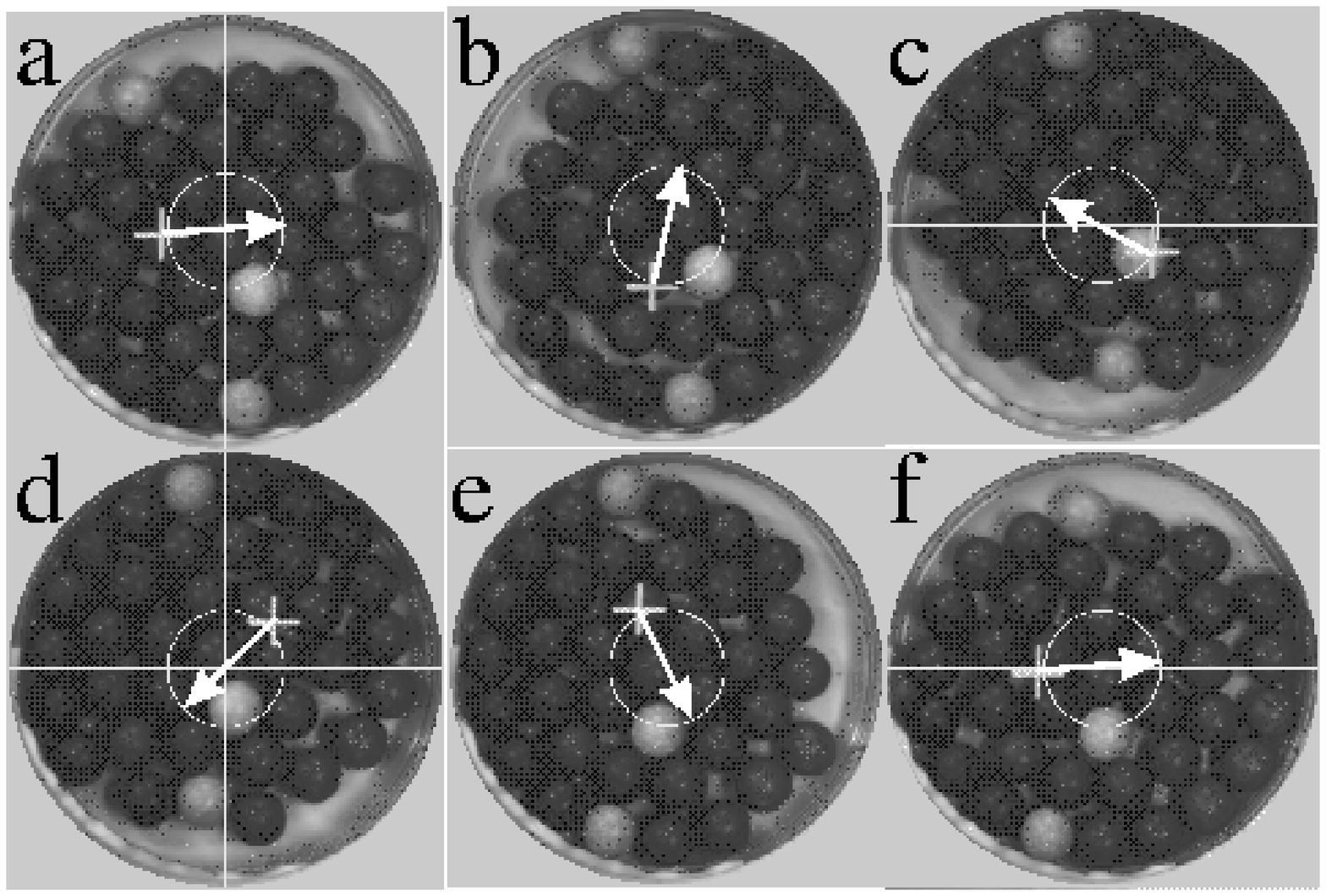,width=7cm,angle=0}}
\psfull
\centerline{\psfig{figure=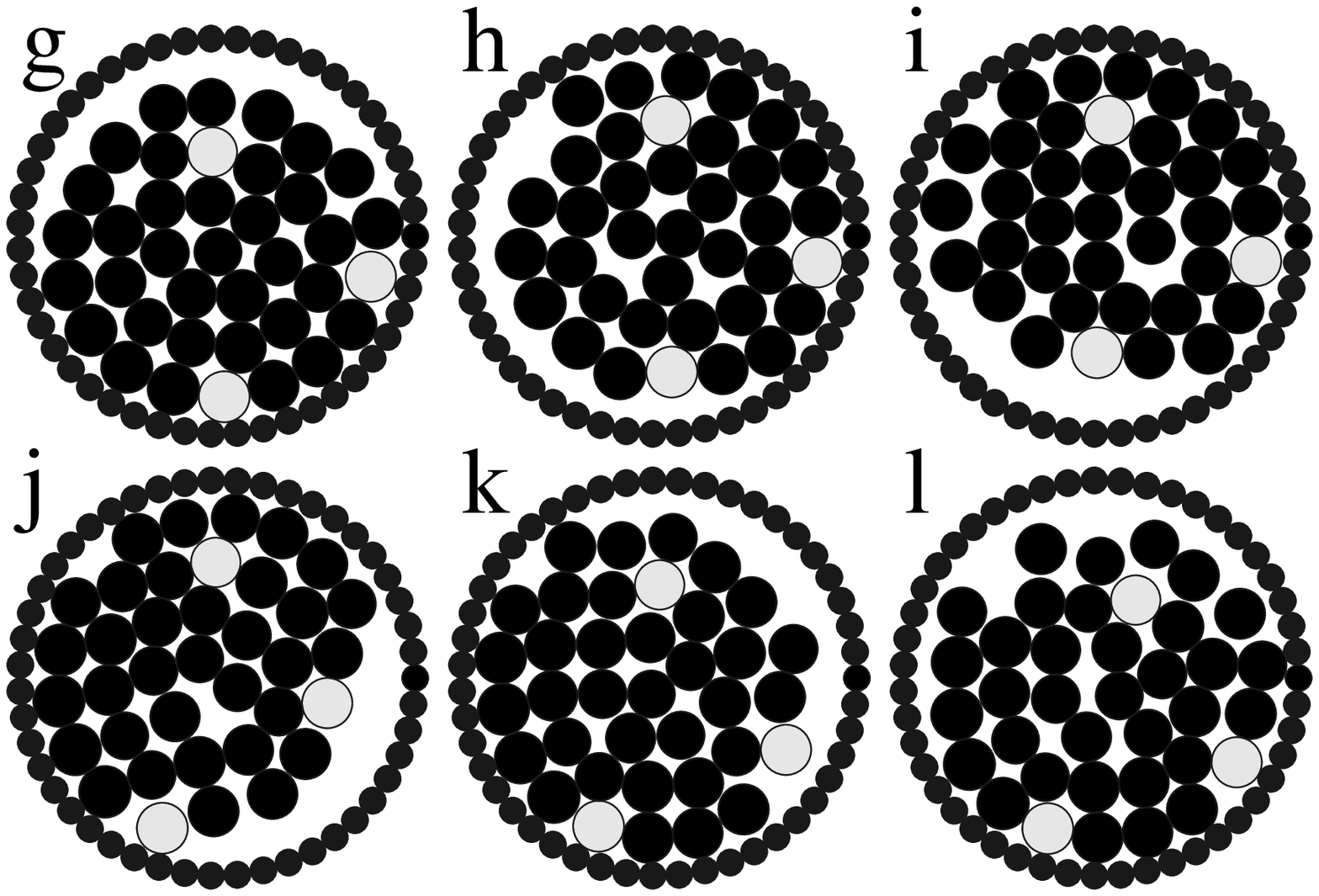,width=7cm,angle=0}}
\caption {Temporal behavior of a system of 37 spheres. Top: the experiment, 
bottom: molecular dynamics simulation.}
\label{MovieHigh}
\end{minipage}
\end{figure}

When the number of spheres is increased the rotation becomes slower.
At a fairly well defined density it stops. Increasing the density then
leads to a different kind of motion, which is displayed in
figs.~\ref{MovieHigh} for the case of 37 spheres. Now the cluster
rotates in the opposite direction. This motion is now better described
by the name reptation. Animated sequences of the motion are available
via WWW~\cite{WWW}.

The experimental snapshots show a white arrow which indicates the direction of
the momentary inertial force. It is obtained by drawing a vector from the
small white cross through the center of the Petri dish. The small white cross
is the marker on the glass plate which is fixed in the laboratory frame as
mentioned above. Therefore it rotates in the comoving frame. This procedure
shows that the phase shift between the crescent-shaped area and the inertial
force is different in the rotation and the reptation mode.

\begin{figure}[htbp]
\begin{minipage}{8 cm}
\centerline{\psfig{figure=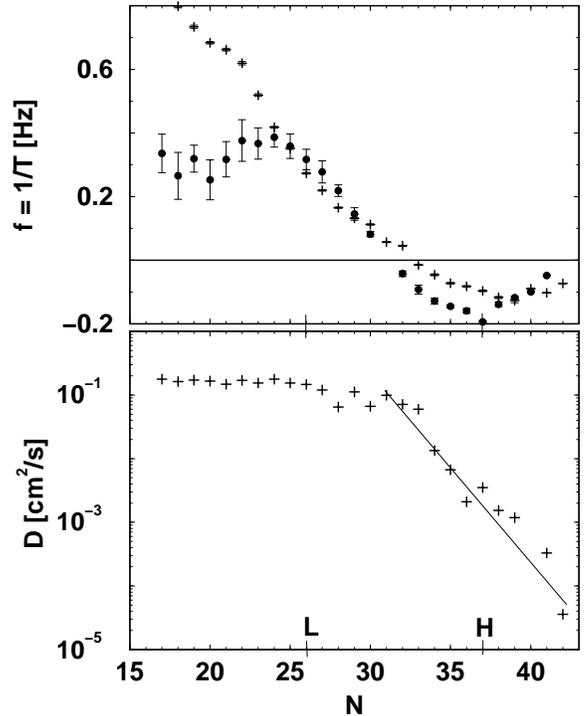,width=7cm,angle=0}}
\caption{The angular velocity of the cluster of particles (top). 
  The solid circles show the experimental data while the crosses show
  the results of a molecular dynamics simulation. For $N_{c}=32$ we observe
the
  reversal of the direction of the rotation. The particle numbers
  according to figures~\ref{MovieLow} and \ref{MovieHigh} have been
  marked by L and H, respectively. The lower part of the figure shows
  the diffusion coefficient $D$ derived from the simulation due to
  eq.~(\ref{diffusion}). For $N$ approaching $N_c$ the diffusion
  coefficient changes significantly.}
\label{hauptbil}
\end{minipage}
\end{figure}

The most striking feature of the motion of the cluster of granular
material is the dependence of the direction and the absolute value of
its angular velocity on the particle number, i.e. the packing density
as presented in the upper part of fig.~\ref{hauptbil}. To obtain the
experimental data indicated by the solid circles, the time for a
single particle located near the wall for one complete path around the
container is measured.  The inverse of this time is shown for
different numbers of spheres.  The effect of the change of the sign of
the rotation is thus clearly demonstrated. This method of measurement
is not sufficiently well defined for a small number of particles. A
particle located at the edge of the cluster does not remain there but
has a tendency to migrate inwards, which makes the characterization of
the cluster movement by tracing an individual particle meaningless.
Therefore in the case of low particle numbers $N<23$ the experimental
method breaks down.

In
order to avoid those difficulties in the simulation we used a slightly different
measurement method for the rotation velocity drawn in fig.~\ref{hauptbil}
(crosses). The motion of the center of
mass is extracted from the numerical data.

\begin{equation}
  \label{measure}
  f= \left< \frac{1}{N} \sum\limits_{i=1}^N \frac{\vec{r}_i^{\,*} \times
      \vec{v}_i^{\,*}}{\left|\vec{r}_i^{\,*}\right|^2} \right>_t~,
\end{equation}
where $\langle \dots \rangle_t$ denotes the time average.
$\vec{r}_i^{\,*}$ and $\vec{v}_i^{\,*}$ are the relative position and velocity
of the $i$th particle with respect to the position and velocity of the
center of mass:
\begin{eqnarray}
  \vec{r}_i^{\,*} &=& \vec{r}_i - \frac{1}{N} \sum\limits_{j=1}^{N} \vec{r}_j \\
  \vec{v}_i^{\,*} &=& \vec{v}_i - \frac{1}{N} \sum\limits_{j=1}^{N} \vec{v}_j~.
\end{eqnarray}

The experiment and the simulation agree quantitatively, showing that
the rotation velocity decreases with particle number. For a critical number of
particles $N_{c}=32$ we observe the reversal of the direction of the rotation.
To obtain the experimental equivalent of eq.~(\ref{measure}) requires computer
aided image analysis which is currently under construction. In the experiment
the rotation velocity is measured from the velocity of particles close to the
wall of the Petri dish only, while in the simulation the velocity is
calculated due to eq.~(\ref{measure}) taking into account data of {\em all}
particles. Hence the particles close to the center of mass having
lower rotation velocity lead to lower values of $f$ for the results of the
simulation (s.~fig.~\ref{hauptbil}). One can get a visual impression of
the described behavior in~\cite{WWW}.

The change of the sense of rotation might be caused by the change of the
dynamical structure of the cluster, which depends on the packing density. For
high densities one finds that the cluster rolls like a rigid body
along the inner wall of the Petri dish, i.e.~the neighborhood
relations of the particles do not change significantly in time. At
lower density the bulk of particles behaves more fluid-like, i.e.~the
relative positions of the particles change quickly in time. A
quantitative measure of this effect is given by the diffusion
coefficient $D$
shown in the lower part of fig.~\ref{hauptbil}:
\begin{equation}
D = \frac{1}{\pi} \left<\frac{d}{dt}\left(\sum\limits_{i=1}^N\sum\limits_{j\in U(i)\ne i}(\vec{r}_i-\vec{r}_j)^2\right)\right>_t .
\label{diffusion}
\end{equation}
Here $U$ denotes the surrounding of the $i$th particle defined by
\begin{equation}
  j\in U(i) \mbox{~~~if~~~} \left| \vec{r}_i-\vec{r}_j \right| \le R_U~.
\end{equation}
In our simulation we used $R_U = 3~cm$. As indicated by 
fig.~\ref{hauptbil}, the diffusion coefficient changes significantly for $N$
approaching $N_c$ . 

The agreement between the experimental results and numerical calculation is
remarkable, when considering the fact that the theory is a two-dimensional one
with idealized boundary conditions and particle interactions. Obviously, this
experiment can serve as a test for the quality of theoretical models
describing the behavior of granular matter, in a field where robust and
reproducible experimental effects are not easily obtained. 

The Deutsche Forschungsgemeinschaft sponsors the experiments through
Re 588/11-1, and the numerical calculations through Ro 548/5-1. It is
a pleasure to thank the ``Abteilung f\"ur Nichtlineare Ph\"anomene''
for enthusiastic support, B.~R\"oder and K.~Scherer for technical
assistance, S.~Seefeld for computer graphics, and A.~Engel, K.~Kassner,
T.~Mahr, St.~Mertens, and L.~Schimansky-Geier for stimulating discussions.

\end{multicols}

\begin{thebibliography}{99}
\bibitem{BlackAndWhite} The black and white spheres have been checked
  to have identical mechanical properties. In test series we did not
  find effects like segregation which might originate from different
  bulk or surface properties.
\end{thebibliography}
\end{document}